\documentclass[prb, aps, reprint, superscriptaddress, floatfix]{revtex4-2}
\usepackage{amsmath}
\usepackage{physics}
\usepackage{graphicx}
\usepackage{amssymb}
\usepackage{amsthm}
\usepackage{mathtools}
\usepackage{braket}
\usepackage[export]{adjustbox}
\usepackage[usenames,dvipsnames]{xcolor}
\definecolor{myblue}{rgb}{0,0,1}
\usepackage[breaklinks=true,colorlinks=true,linkcolor=myblue,citecolor=myblue]{hyperref}

\newcommand{\omax}{\omega_{\max}}

\newcommand{\OO}[1]{\mathcal{O}(#1)}
\newcommand{\paren}[1]{\left( #1 \right)}

\newcommand{\wh}[1]{\widehat{#1}}
\newcommand{\wt}[1]{\widetilde{#1}}

\newcommand{\nmax}{n_{\max}}

\newcommand{\CCQ}{Center for Computational Quantum Physics, Flatiron Institute, 162 5th Avenue, New York, NY 10010, USA}
\newcommand{\CCM}{Center for Computational Mathematics, Flatiron Institute, 162 5th Avenue, New York, NY 10010, USA}

\newtheorem{remark}{Remark}

\begin{document}

\title{Discrete Lehmann representation of three-point functions}

\author{Dominik Kiese}
\affiliation{\CCQ}

\author{Hugo U. R. Strand}
\affiliation{School of Science and Technology, Örebro University, SE-701 82 Örebro, Sweden}

\author{Kun Chen}
\affiliation{CAS Key Laboratory of Theoretical Physics, Institute of Theoretical Physics, Chinese Academy of Sciences, Beijing 100190, China}
\affiliation{\CCQ}

\author{Nils Wentzell}
\affiliation{\CCQ}

\author{Olivier Parcollet}
\affiliation{\CCQ}
\affiliation{Universit\'e Paris-Saclay, CNRS, CEA, Institut de Physique Th\'eorique, 91191, Gif-sur-Yvette, France}

\author{Jason Kaye}
\email{jkaye@flatironinstitute.org}
\affiliation{\CCQ}
\affiliation{\CCM}

\begin{abstract}
	We present a generalization of the discrete Lehmann representation (DLR)
	to three-point correlation and vertex functions in imaginary time and Matsubara frequency.
	The representation takes the form of a
	linear combination of judiciously chosen exponentials in imaginary time, and
	products of simple poles in Matsubara frequency, which are universal for a
	given temperature and energy cutoff. We present a systematic algorithm to
	generate compact sampling grids, from which the coefficients of such an
	expansion can be obtained by solving a linear system. We show that the
	explicit form of the representation can be used to evaluate diagrammatic expressions involving infinite Matsubara
	sums, such as polarization functions or self-energies, with controllable,
	high-order accuracy. This collection of techniques establishes a framework
	through which methods involving three-point objects can be implemented
	robustly, with a substantially reduced computational cost and memory footprint.
\end{abstract}

\maketitle

\section{Introduction}

One and two-body correlation functions
play a central role in theoretical condensed matter physics, 
in particular in the study of strong electronic correlation effects in materials and devices.
They are directly related to experimental quantities, such as transport via the Kubo formula.
They are also the natural language used in diagrammatic techniques and some computational methods, including 
dynamical mean-field theory (DMFT) \cite{Georges_1996, Kotliar_2006}
and its extensions \cite{Rohringer_2018}.
Their efficient representation is therefore of great practical importance in computational quantum physics.
The challenge consists in obtaining compressed representations for these functions
(to decrease their memory footprint) in a manner compatible with efficient 
algorithms for standard operations.
Among those correlations functions, the three-point vertex already constitutes
a building block for several non-perturbative approaches
\cite{Ayral2015, Ayral_2016, Rohringer_2018, Krien_2020, Krien_2021}. 

A typical standard discretization of a correlation function in the imaginary
frequency domain consists of its values on a product grid containing all
tuples of Matsubara frequencies up to a given high-frequency cutoff
\cite{vanloon23, Rohringer_2018, Stepanov_2019, Ayral_2016, Wentzell_2020}. Such an
approach is low-order accurate and memory intensive, particularly because of the slow decay of
these functions to their asymptotic limit. 
Furthermore, diagrammatic expressions involving infinite
Matsubara frequency sums are typically computed by brute force, which is
prohibitively expensive for expressions involving even two or three frequency
variables \cite{vanloon23, Rohringer_2018, Stepanov_2019, Ayral_2016, Wentzell_2020, chen2024partial}. The significant memory associated with storing these objects has motivated the development of compression
techniques based on neural networks, principal components analysis, and wavelets
\cite{disante22, zang24, moghadas24}, but without efficient algorithms to perform standard operations. A representation in terms of the orthogonal Koornwinder-Dubiner polynomials \cite{koornwinder75,dubiner91,sherwin95} was used for Monte Carlo sampling on the imaginary time simplex in Ref.~\onlinecite{kim22}, but this approach does not make use of the specific structure of the underlying objects and therefore yields suboptimal compression. A compression method based on
quantics tensor trains has recently been proposed \cite{shinaoka23,takahashi24}, but further work is required to explore the properties of
this representation and its potential for practical algorithms.

Recently, significant progress has been made on the efficient discretization of single-particle imaginary time/frequency Green's functions and
other single-variable quantities, such as self-energies and hybridization
functions, using low-rank compression techniques based on spectral
representations. This began with the intermediate representation (IR), an
orthogonal basis spanning the set of imaginary time Green's functions with remarkably
few degrees of freedom \cite{shinaoka17,chikano18}. The discrete Lehmann
representation (DLR) \cite{kaye22_dlr} spans the same set using simple, explicit basis functions,
given by exponentials in imaginary time and simple poles in Matsubara frequency. Such representations offer user-controllable accuracy up to
machine precision, efficient mapping between grid and coefficient
representations \cite{li20,kaye22_dlr}, and in the case of the DLR, explicit formulas for standard
operations such as Fourier transform, convolution, and inner product.
They have entered routine use in many-body calculations, in many cases
replacing generic discretizations such as uniform grids and orthogonal polynomials
\cite{boehnke11, gull18, dong20}, and providing the basis of several algorithmic developments in DMFT \cite{sheng23,labollita23}, diagrammatic methods \cite{kaye23_diagrams},
calculations on the Keldysh contour \cite{kaye23_eqdyson,blommel24}, and self-consistent Dyson equation solvers \cite{yeh22, cai2022, wang2023}. Several open-source implementations are available
\cite{chikano19,kaye22_libdlr,wallerberger23,kaye24_cppdlr,parcollet15}.

Ref.~\onlinecite{shinaoka18} introduced
a spectral representation of two-particle Green's functions, similar
to that for the single-particle case, and used it to build
\emph{overcomplete} yet highly compact representations in terms of the IR basis
functions. Ref.~\onlinecite{shinaoka20} then developed compact sampling
grids from which such expansions can be formed, and Ref.~\onlinecite{wallerberger21} used these tools to
solve the Bethe-Salpeter equation (BSE) for several model problems. This
approach shares the primary advantages of the single-particle case: it is
compact even at low temperatures, general, and offers user-controllable,
high-order accuracy.

In this work, we follow a similar path to
Refs.~\onlinecite{shinaoka18,shinaoka20,wallerberger21}, but using the DLR rather than the IR basis. 
This approach yields an explicit representation
of correlation functions in terms of products of simple poles in Matsubara
frequency and exponentials in imaginary time. Using the representation, we obtain efficient
algorithms to compute the infinite
Matsubara frequency sums appearing in standard diagrammatic expressions,
facilitating its use in practical many-body calculations. In
particular, we present two techniques based on our expansion: one using
separation of variables, and another exploiting its explicit form using residue calculus. In
addition, we obtain a more compact grid representation than the overcomplete
sparse sampling approach used in Refs.~\onlinecite{shinaoka20,wallerberger21}, and suggest a
simple interpolation method to obtain the expansion coefficients without the
need for regularization parameters. Even though we focus on three-point correlation
and vertex functions in this work, the techniques we present
generalize straightforwardly to the four-point case. We demonstrate the
accuracy of the representation for the three-point correlation and
vertex functions of both the Hubbard atom and an interacting impurity coupled to a few bath sites.
We then apply our Matsubara summation algorithm to compute the polarization functions in each case.
The
collection of tools we present provides a framework for robust and efficient
computations involving three-point functions in quantum many-body physics calculations.

\section{Background and methodology}

\subsection{Discrete Lehmann representation of single-particle Green's functions} \label{sec:dlr1d}

The DLR is based on the spectral Lehmann representation, an
integral representation satisfied by all single-particle Matsubara Green's
functions \cite{kaye22_dlr}. For fermionic Green's functions, it can be written
\begin{equation} \label{eq:lehmann1d}
    G(i \nu_n) = \int_{-\infty}^{\infty} d\omega \, K(i \nu_n, \omega) \rho(\omega),
\end{equation}
where $i \nu_n = (2n + 1) \pi i / \beta$ are the fermionic Matsubara frequencies,
$\rho$ is an absolutely integrable spectral density, and
\begin{equation} \label{eq:ackernel}
    K(i \nu_n, \omega) = (i \nu_n - \omega)^{-1}
\end{equation}
is the analytic continuation kernel. For bosonic Green's functions, we use the
representation \eqref{eq:lehmann1d} with $K$ replaced by
\begin{equation}
    K_B(i \Omega_n, \omega) = \frac{\tanh(\beta \omega / 2)}{i \Omega_n - \omega},
\end{equation}
for $i \Omega_n = 2 \pi n i / \beta$ the bosonic Matsubara frequencies.
An analogous integral representation holds in imaginary time. It is convenient to work in
dimensionless variables $\nu_n \gets \beta \nu_n$, $\omega \gets \beta \omega$,
for which, assuming $\rho = 0$ outside $[-\omax, \omax]$, we obtain
\begin{equation} \label{eq:lehmanntrunc}
    G(i \nu_n) = \int_{-\Lambda}^{\Lambda} d\omega \, K(i \nu_n, \omega) \rho(\omega)
\end{equation}
with rescaled $G$, $\rho$. Here $\Lambda = \beta \omax$ is a dimensionless
high-frequency cutoff parameter.

The integral operator in \eqref{eq:lehmanntrunc} has
super-exponentially decaying singular values \cite{shinaoka17,chikano18}, and an
explicit low-rank approximation of its kernel can be obtained
via the interpolative decomposition (ID) \cite{kaye22_dlr,cheng05,liberty07}:
\begin{equation} \label{eq:dlrkernel}
    K(i \nu_n, \omega) \approx \sum_{l=1}^r K(i \nu_n, \omega_l) \pi_l(\omega).
\end{equation}
Here, the ID, implemented via the rank-revealing pivoted Gram-Schmidt algorithm, selects $r$
frequencies $\omega_l$, called the DLR frequencies, such that an approximation
of the form \eqref{eq:dlrkernel} can be recovered in a numerically stable
manner, as described below.
The accuracy of this approximation can be specified and guaranteed by an
appropriate truncation of the pivoting procedure, so that the kernel can be approximated to
$\epsilon$ accuracy by $r = \OO{\log(\Lambda) \log(\epsilon^{-1})}$ simple
poles. Substitution of \eqref{eq:dlrkernel} into \eqref{eq:lehmanntrunc} yields
the DLR of $G$,
\begin{equation} \label{eq:dlr}
    G(i \nu_n) \approx \sum_{l=1}^r K(i \nu_n, \omega_l) \wh{g_l}.
\end{equation}
Here, the DLR coefficients $\wh{g_l}$ can be computed directly if $\rho$ is
known, but typically this is not the case. However, by performing a further ID on the Matsubara frequency variable of the
remaining kernel $K(i \nu_n, \omega_l)$, one can obtain a collection of $r$ DLR interpolation nodes
$i \nu_{n_k}$ such that the DLR coefficients are the solution of the following interpolation problem:
\begin{equation} \label{eq:dlrinterp}
    \sum_{l=1}^r K(i \nu_{n_k}, \omega_l) \wh{g_l} = G(i \nu_{n_k}).
\end{equation}
The LU factors of the $r \times r$ matrix $K(i \nu_{n_k}, \omega_l)$ can be precomputed
and stored, so solving \eqref{eq:dlrinterp} to obtain the DLR coefficients has
an $\OO{r^2}$ computational cost.
One can similarly obtain a collection of $r$ bosonic interpolation nodes $i
\Omega_{n_k^B}$, with an analogous interpolation problem for bosonic Green's functions.

Fourier transform of \eqref{eq:dlr} yields
a DLR expansion in imaginary time,
\begin{equation} \label{eq:dlrtime}
    G(\tau) \approx \sum_{l=1}^r K(\tau, \omega_l) \wh{g_l},
\end{equation}
with
\begin{equation}
    K(\tau, \omega) = -\frac{e^{-\omega \tau}}{1 + e^{-\beta \omega}}
\end{equation}
for $\tau \in [0, \beta]$, given here in the original dimensional variables,
for both the fermionic and bosonic kernels.
Imaginary time interpolation nodes $\tau_k$ can be similarly constructed. In general, given
user-specified $\Lambda$ and $\epsilon$, the DLR
frequencies $\omega_l$ may be selected so as to guarantee the approximation
error $\norm{G - G_\text{DLR}}_2 < \epsilon$ for any $G$ with spectral width
$\omax$, with $G_\text{DLR}$ the DLR expansion and the $L^2(\tau)$ and $l^2(i \nu_n)$ norms given by \cite{wang24}
\begin{equation} \label{eq:l2norm}
    \norm{G}_2^2 = \frac{1}{\beta} \int_0^\beta d\tau \, \abs{G(\tau)}^2 = \frac{1}{\beta^2} \sum_{n=-\infty}^\infty \abs{G(i \nu_n)}^2.
\end{equation}

We summarize three key properties of the DLR:
\begin{enumerate}
    \item The DLR basis functions $K(i \nu_n, \omega_l)$ are universal given a
    specified dimensionless frequency cutoff $\Lambda$ and accuracy $\epsilon$;
    only the coefficients $\wh{g_l}$ depend on the Green's function, not the DLR
    frequencies $\omega_l$.
    \item The DLR has only $r = \OO{\log(\Lambda)
    \log(\epsilon^{-1})}$ degrees of freedom, and is therefore significantly
    more compact than uniform grid ($\OO{\Lambda}$ degrees of freedom with low-order
    accuracy) or orthogonal polynomial ($\OO{\sqrt{\Lambda}}$
    degrees of freedom) \cite{boehnke11, gull18, dong20} representations.
    \item The DLR basis functions are explicitly known, so explicit formulas for
    operations such as Fourier transform \cite{kaye22_dlr}, imaginary time
    convolution \cite{kaye23_diagrams}, and
    Matsubara frequency summation can be obtained by applying them to the basis
    functions.
\end{enumerate}

\subsection{Discrete Lehmann representation of three-point correlation functions}

We consider the following three-point imaginary time correlation functions in the particle-particle ($pp$) and particle-hole ($ph$) channels:
\begin{equation}
    \begin{aligned}
        \chi^{pp}(\tau_1, \tau_2) &= \braket{\hat{T} c(\tau_1) c(\tau_2) \Delta^{\dagger}(0)}, \\ 
        \chi^{ph}(\tau_1, \tau_2) &= \braket{\hat{T} c^{\dagger}(\tau_1) c(\tau_2) n(0)}.
    \end{aligned}
\end{equation}
Here $\hat{T}$ is the imaginary time-ordering operator, $\Delta^{\dagger} =
c^{\dagger} c^{\dagger}$ represents pairing fluctuations, $n =
c^{\dagger} c$ density fluctuations, and for simplicity we
have suppressed spatial and orbital degrees of freedom. We define the correlation
functions in Matsubara frequency by
\begin{equation}
   \chi^{\xi}(i \nu_m, i \nu_n) = \int_{[0,\beta]^2} d\tau_1 d\tau_2 \, e^{-i(\xi \nu_m \tau_1 + \nu_n \tau_2)} \chi^{pp}(\tau_1, \tau_2).
\end{equation}
We write $\xi = \pm 1$ for the particle-particle and
particle-hole channels, respectively, and as a shorthand write $\chi^\xi$ to
refer to $\chi^{pp}$ or $\chi^{ph}$. 

The three-point correlation functions obey the spectral representation
\begin{equation} \label{eq:specrep_mf}
    \begin{multlined} 
        \chi^\xi(i \nu_m, i \nu_n) 
        \\ = \int_{-\infty}^\infty \int_{-\infty}^\infty d\omega_1 d\omega_2 \, \Big(K(i \xi \nu_m, \omega_1) K(i \nu_n, \omega_2) \rho_1(\omega_1, \omega_2) \\ 
        + K(i \nu_n, \omega_2) K_B(i \xi \nu_m + i\nu_n, \omega_1 + \omega_2) \rho_2(\omega_1, \omega_2) \\
        + K(i \xi \nu_m, \omega_1) K_B(i \xi \nu_m + i \nu_n, \omega_1 + \omega_2) \rho_3(\omega_1, \omega_2) \Big) \\
        + \beta \delta_{\xi \nu_m + \nu_n, 0} \int_{-\infty}^\infty d\omega \, K(i\xi \nu_m, \omega) \rho_4(\omega),
    \end{multlined}
\end{equation}
with spectral densities $\rho_1, \ldots, \rho_4$. The first three
terms comprise the \emph{regular} part of the representation, and the final term
the \emph{singular} part. We refer the reader to
Appendix B of Ref.~\onlinecite{shinaoka18} for a detailed derivation of this
representation, and to Refs.~\onlinecite{kugler21,dirnbock24} for further discussion of spectral representations of $n$-point correlation functions.
Our representation differs from that of
Ref.~\onlinecite{shinaoka18} only in that we explicitly distinguish between the
particle-particle and particle-hole channels.
As in \eqref{eq:lehmanntrunc}, \eqref{eq:specrep_mf} can be truncated and
non-dimensionalized, yielding an analogous expression with
the domain of integration replaced by $[-\Lambda,\Lambda]^2$ for the regular part and
$[-\Lambda, \Lambda]$ for the singular part.

As in the derivation of \eqref{eq:dlr}, we
substitute \eqref{eq:dlrkernel} into the non-dimensionalized spectral representation, yielding
\begin{equation} \label{eq:dlr2d}
    \begin{aligned}
        \chi^\xi(i \nu_m, i \nu_n)
        &\begin{multlined}[t]
        \approx \sum_{k,l=1}^r \Big( K(i\xi \nu_m, \omega_k) K(i\nu_n, \omega_l) \wh{\chi}_{1kl} \\
        + K(i\nu_n, \omega_k) K_B(i\xi \nu_m + i \nu_n, \omega_l) \wh{\chi}_{2kl} \\
        + K(i\xi \nu_m, \omega_k) K_B(i\xi \nu_m + i\nu_n, \omega_l) \wh{\chi}_{3kl} \Big) \\
        + \delta_{\xi \nu_m + \nu_n, 0} \sum_{k=1}^r K(i\xi \nu_m, \omega_k) \wh{\chi}_{4kk}
        \end{multlined} \\
        &= \sum_{\alpha=1}^4 \sum_{k,l=1}^r \phi_{\alpha kl}^\xi(i \nu_m,i \nu_n) \wh{\chi}_{\alpha kl},
\end{aligned}
\end{equation}
with
\begin{equation} \label{eq:dlr2dcoefs}
    \begin{aligned}
        \wh{\chi}_{1kl} &= \int_{[-\Lambda,\Lambda]^2} d\omega_1 d\omega_2 \, \pi_k(\omega_1) \pi_l(\omega_2) \rho_1(\omega_1, \omega_2), \\
        \wh{\chi}_{2kl} &= \int_{[-\Lambda,\Lambda]^2} d\omega_1 d\omega_2 \, \pi_k(\omega_2) \pi_l(\omega_1 + \omega_2) \rho_2(\omega_1, \omega_2), \\
        \wh{\chi}_{3kl} &= \int_{[-\Lambda,\Lambda]^2} d\omega_1 d\omega_2 \, \pi_k(\omega_1) \pi_l(\omega_1 + \omega_2) \rho_3(\omega_1, \omega_2), \\
        \wh{\chi}_{4kk} &= \beta \int_{-\Lambda}^\Lambda d\omega \, \pi_k(\omega) \rho_4(\omega),
    \end{aligned}
\end{equation}
and
\begin{equation} \label{eq:dlrbasis}
    \begin{aligned}
        \phi_{1kl}^\xi(i \nu_m, i \nu_n) &= K(i\xi \nu_m, \omega_k) K(i\nu_n, \omega_l), \\
        \phi_{2kl}^\xi(i \nu_m, i \nu_n) &= K(i\nu_n, \omega_k) K_B(i\xi \nu_m + i\nu_n, \omega_l), \\
        \phi_{3kl}^\xi(i \nu_m, i \nu_n) &= K(i\xi \nu_m, \omega_k) K_B(i\xi \nu_m + \nu_n, \omega_l), \\
        \phi_{4kl}^\xi(i \nu_m, i \nu_n) &= \delta_{\xi \nu_m + \nu_n, 0} \delta_{kl} K(i\xi \nu_m, \omega_k).
    \end{aligned}
\end{equation}
As before, the densities $\rho_\alpha$, assumed to be supported in $[-\Lambda, \Lambda]^2$, are typically not known, so the coefficients
\eqref{eq:dlr2dcoefs} cannot be computed directly and \eqref{eq:dlr2d}
simply establishes the existence of such a representation.
We consider \eqref{eq:dlr2d} to be a two-dimensional generalization of the DLR.
Given the cutoff $\Lambda$, and the error tolerance $\epsilon$, the functions
$\phi_{\alpha kl}^\xi$ are again independent of $\chi^\xi$ itself, and the
number of degrees
of freedom in the representation is $3r^2 + r = \OO{\log^2(\Lambda)
\log^2(\epsilon^{-1})}$. We see from \eqref{eq:dlrbasis} that the $\phi_{\alpha kl}^\xi$ are 
simple rational functions.

Although the $\phi_{\alpha kl}^\xi$ span the subspace
determined by the representation \eqref{eq:specrep_mf}, we will see that they are not numerically linearly independent. This is the origin of the terminology ``overcomplete
representation'' used for the similar expansion obtained using the IR basis in
Ref.~\onlinecite{shinaoka18}.

Fourier transform of \eqref{eq:dlr2d} gives an analogous imaginary time DLR
expansion,
\begin{equation} \label{eq:dlr2dtime}
    \begin{aligned}
        \chi^\xi(\tau_1, \tau_2)
        &\approx \sum_{k,l=1}^r \Big( K(\tau_1, \omega_k) K(\tau_2, \omega_l) \wh{\chi}_{1kl} \\
        &+ K(\tau_2-\tau_1, \omega_k) K(\tau_1, \omega_l) \wh{\chi}_{2kl} \\
        &+ K(\tau_1-\tau_2, \omega_k) K(\tau_2, \omega_l) \wh{\chi}_{3kl} \Big) \\
        &+ \sum_{k=1}^r K(\tau_1-\tau_2, \omega_k) \wh{\chi}_{4kk}
    \end{aligned}
\end{equation}
so that a DLR expansion obtained in Matsubara frequency can be evaluated in
imaginary time, and vice versa. Here, $\tau_1, \tau_2 \in (0, \beta)$, and
we extend $K(\tau, \omega)$ to $(-\beta, 0)$ using the anti-periodicity
condition $K(\tau, \omega) = -K(\beta + \tau, \omega) = -K(-\tau, -\omega)$ for
$\tau \in (-\beta, 0)$.

\subsection{Matsubara frequency interpolation} \label{sec:dlrinterp}

To obtain an expansion \eqref{eq:dlr2d} from evaluations of $\chi^\xi$ in the
Matsubara frequency domain, we follow a similar strategy to that used in
Ref.~\onlinecite{kaye22_dlr} to obtain the DLR imaginary time and frequency
grids. For simplicity, we fix $\xi = pp$ and suppress the $\xi$ index (see
Remark \ref{rem:phpp} below).
Introducing a Matsubara frequency cutoff $\nmax$, we form the $(2 \nmax)^2 \times (3r^2 + r)$ matrix 
\begin{equation} \label{eq:finesysmat}
    \Phi^f_{mn,\alpha kl} = \phi_{\alpha kl}(i \nu_m, i \nu_n),
\end{equation}
with $m,n \in \{-\nmax, \ldots, \nmax - 1\}$, $\alpha \in \{1,\ldots,4\}$, and $k, l \in
\{1,\ldots, r\}$ (entries with $\alpha = 4$ and $k \neq l$ are zero and not included). Here, the indices $m$, $n$ and $\alpha$, $k$,
$l$ are grouped together to obtain the matrix representation, and the
superscript ``$f$'' indicates that we have sampled on a \emph{fine} or uncompressed
grid. We can then carry out the rank-revealing pivoted
Gram-Schmidt procedure
on the rows of $\Phi^f$, until the user-specified tolerance $\epsilon$ is
reached \cite{cheng05}.
The selected pivot rows correspond to a collection of \emph{DLR nodes} $(i \nu_{m_j}, i
\nu_{n_j})$, $j \in \{1,\ldots,R\}$, a subset of Matsubara frequency points which form a well-conditioned interpolation
grid. Here, $R$ is an estimate of the $\epsilon$-rank of $\Phi^f$.
We can converge this subsampling procedure with respect to the choice of $\nmax$ such that the selected nodes
no longer change, which typically occurs for $\nmax \propto \Lambda$.

\begin{figure}
    \includegraphics[width=.85\columnwidth]{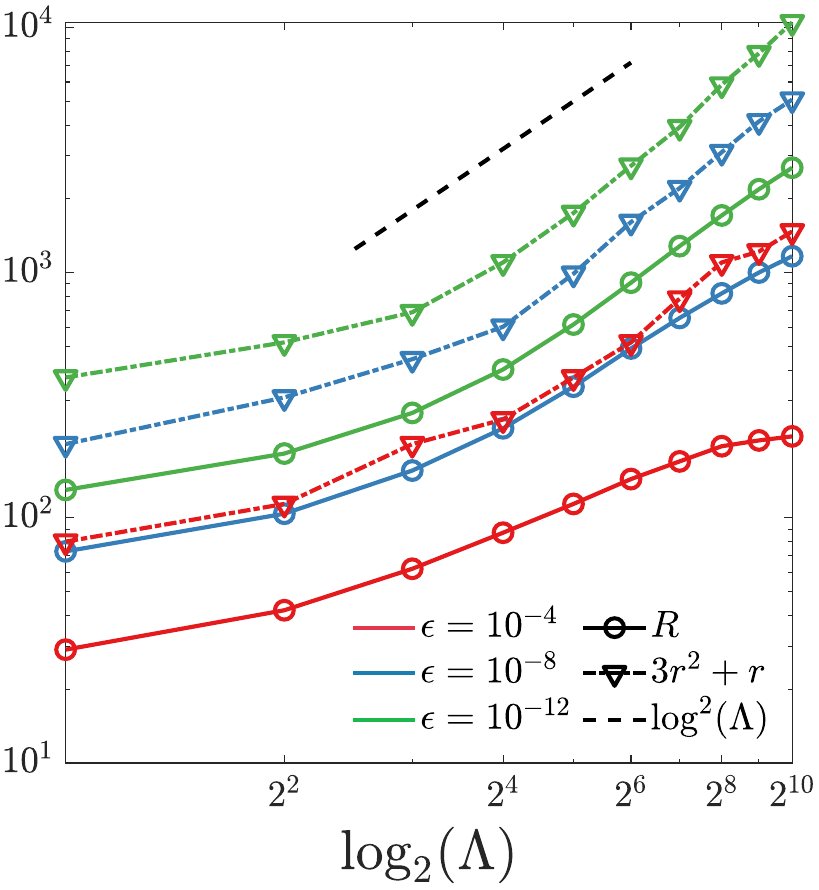}
    \caption{$\epsilon$-rank $R$ of $\Phi^f$ (solid lines), which determines the number of DLR grid
    points, or degrees of freedom in our representation, as a function of
    $\Lambda$ and $\epsilon$. The quantity $3r^2 + r$ (dash-dotted lines), the maximum possible
    rank, is also plotted. $R$ scales as $\OO{\log^2(\Lambda)}$, and is less
    than the maximum rank, indicating that further compression of the
    representation \eqref{eq:dlr2d} is possible.}
    \label{fig:rankvlambda}
\end{figure}

We next define the
$R \times (3r^2 + r)$ matrix
\begin{equation}
    \Phi_{j,\alpha kl} = \phi_{\alpha kl}(i \nu_{m_j}, i \nu_{n_j}).
\end{equation}
Then, given the values of $\chi$ at the DLR nodes, we can obtain the
DLR coefficients $\wh{\chi}_{\alpha kl}$ by solving the rectangular linear system
\begin{equation} \label{eq:dlr2dlin}
    \sum_{\alpha=1}^4 \sum_{k,l=1}^r \Phi_{j, \alpha k l} \, \wh{\chi}_{\alpha k l} = \chi(i \nu_{m_j}, i \nu_{n_j}).
\end{equation}
Fig. \ref{fig:rankvlambda} shows $R$ and $3r^2 + r$ as functions of $\Lambda$ for
several choices of $\epsilon$. We find $R < 3r^2 + r$, indicating that
$\Phi^f$ is rank-deficient. This implies that \eqref{eq:dlr2dlin} is
underdetermined, and we solve it in the minimum norm sense using the
pseudoinverse. We expect $R = \OO{r^2} = \OO{\log^2(\Lambda)}$, and find this to
be the case, except for slight deviations at larger values of $\Lambda$; these
might result from the choice of truncation criteria in the
rank-revealing pivoted Gram-Schmidt procedure, which should be studied carefully
in future work.

\begin{remark}
    The $\epsilon$-rank deficiency of $\Phi^f$ implies that the functions $\phi_{\alpha kl}$ are numerically linearly
    dependent, and could be compressed into a basis of $R$ functions by a similar
    pivoted Gram-Schmidt procedure on the columns of $\Phi^f$ or $\Phi$ (as is
    done to obtain the one-dimensional DLR basis). Although such a recompression
    procedure leads to a mild reduction in the number of terms in the
    expansion, it does not reduce the overall $\OO{r^2}$ scaling, and we find that
    the solution of the underdetermined system \eqref{eq:dlr2dlin} via the
    pseudoinverse yields slightly better accuracy than the solution of the square system 
    obtained by recompressing the $\phi_{\alpha kl}$. Since this recompression does not lead to
    a critical performance improvement for the examples we consider, we defer
    its exploration to future studies; for instance, for four-point functions,
    such a recompression may be helpful.
\end{remark}

\begin{remark} \label{rem:phpp}
    If a function $\chi^{ph}$ obeys the particle-hole representation in
    \eqref{eq:dlr2d}, then $\chi^{ph}(-i \nu_m, i \nu_n)$ obeys the
    particle-particle representation. Therefore, if $(i \nu_{m_j}, i \nu_{n_j})$
    are the DLR nodes for the
    particle-particle representation, then the DLR nodes for the particle-hole
    representation are simply $(-i \nu_{m_j}, i \nu_{n_j})$; in other words, the
    DLR coefficients of $\chi^{ph}$ can be obtained by solving
    \eqref{eq:dlr2dlin} with the right hand side replaced by $\chi^{ph}(-i
    \nu_{m_j}, i \nu_{n_j})$.
\end{remark}

\begin{figure}
    \includegraphics[width=\columnwidth]{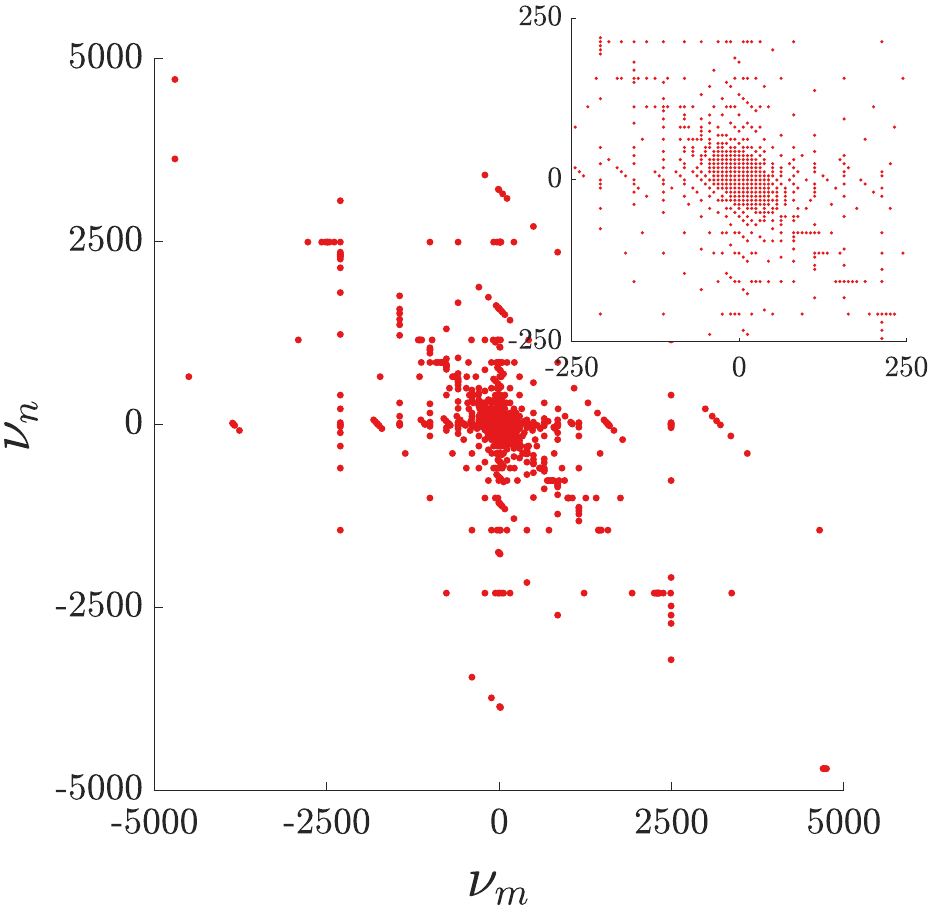}
    \caption{The $R = 1180$ DLR nodes $(i\nu_{m_j}, i\nu_{n_j})$ for $\Lambda = 1024$ and
    $\epsilon = 10^{-8}$, in dimensionless units. The inset is a zoom-in to the
    low-frequency region.}
    \label{fig:dlrnodes}
\end{figure}

An example of the DLR nodes obtained for $\Lambda = 1024$ and
$\epsilon = 10^{-8}$ is shown in Fig.
\ref{fig:dlrnodes}. The selected nodes reflect the structure of typical
three-point correlation functions, as determined by the spectral representation
\eqref{eq:specrep_mf}.

\subsection{Reduced fine Matsubara frequency grid}
In obtaining the DLR nodes, we must take $\nmax = \OO{\Lambda}$ to ensure a sufficiently fine sampling of
Matsubara frequencies. Thus, the number of rows of $\Phi^f$ scales as $\OO{\Lambda^2}$,
whereas the number of columns scales as $\OO{r^2} = \OO{\log^2(\Lambda)}$.
Although the DLR nodes can be precomputed and stored with a
negligible memory footprint, the cost of this precomputation scales as
$\OO{\nmax^2 r^2 R} = \OO{\Lambda^2 \log^4(\Lambda)}$ (since the cost of
rank-revealing pivoted Gram-Schmidt on an $m \times n$ matrix of rank $k$ is $\OO{mnk}$). We reduce this cost
using a more efficient fine Matsubara frequency grid, defined as the union of
the following collections of grid points:
\begin{equation} \label{eq:finegrid}
    \begin{aligned}
        (\nu^f_{1p}, \nu^f_{1q}) & = (\nu_{\bar{n}_p}, \nu_{\bar{n}_q}), \\
        (\nu^f_{2p}, \nu^f_{2q}) & = (\Omega_{\bar{n}_p^B} - \nu_{\bar{n}_q}, \nu_{\bar{n}_q}), \\
        (\nu^f_{3p}, \nu^f_{3q}) & = (\nu_{\bar{n}_p}, \Omega_{\bar{n}_q^B} - \nu_{\bar{n}_p}), \\
        (\nu^f_{4p}, \nu^f_{4p}) & = (\nu_{\bar{n}_p}, -\nu_{\bar{n}_p}).
    \end{aligned}
\end{equation}
Here, $p, q \in \{1,\ldots, r\}$, yielding a total of $3r^2 + r$ nodes.
$\nu_{\bar{n}_p}$ and $\Omega_{\bar{n}_p^B}$ are the one-dimensional fermionic and bosonic
DLR nodes, respectively, discussed in Sec.
\ref{sec:dlr1d}, with the $\bar{\cdot}$ notation distinguishing them from the two-dimensional DLR nodes.
This choice
arises as the union of the Cartesian products of one-dimensional DLR grids
corresponding to the first three terms of \eqref{eq:dlr2d}, together with the
grid points along the off-diagonal corresponding to the singular term. It has
the effect of clustering grid points according to the features of the functions $\phi_{\alpha kl}$.

To obtain the DLR nodes, we simply replace $\Phi^f$ defined by
\eqref{eq:finesysmat} with the $(3r^2 + r) \times (3r^2 + r)$ matrix
\begin{equation}
    \Phi^f_{\gamma pq, \alpha kl} = \phi_{\alpha kl}(i \nu^f_{\gamma p}, i \nu^f_{\gamma q}),
\end{equation}
for $\alpha, \gamma \in \{1,\ldots,4\}$ and $k,l,p,q \in \{1,\ldots,r\}$ (entries with
$\alpha = 4$ and $k \neq l$ or $\gamma = 4$ and $p \neq q$ are not included),
in the procedure described above. This simply amounts to subselecting from a
more compact fine grid of Matsubara frequency pairs.

In practice, we find that this approach produces grids of equally good quality,
while reducing the cost of precomputing the DLR nodes to $\OO{r^4 R} = \OO{r^6} =
\OO{\log^6 \Lambda}$, typically modest in practice. For example, for $\Lambda =
1000$ and $\epsilon = 10^{-8}$ we have $r = 42$, so $\Phi^f$ is of size $5334 \times
5334$ and the Gram-Schmidt procedure can be carried out on a laptop in
approximately $30$ seconds. We note that except for the choice of the underlying
one-dimensional grids, the grid points \eqref{eq:finegrid} are
the same as the \emph{sparse sampling} grid used in
Ref.~\onlinecite{wallerberger21}; however, whereas we further compress the fine grid
to a grid of $R$ DLR nodes, Ref.~\onlinecite{wallerberger21} uses the sparse sampling
grid points without further compression to compute the expansion coefficients of the
overcomplete representation.

\section{Matsubara frequency summation} \label{sec:matsum}

The simple form of the DLR expansion \eqref{eq:dlr2d} leads to
efficient algorithms for computing the infinite Matsubara frequency sums
appearing in diagrammatic expressions. As an example, we focus on the
polarization functions
\begin{widetext}
\begin{equation} \label{eq:polppph}
    \begin{aligned}
        P^{pp}(i \Omega_m) &= -\frac{1}{2\beta} \sum_{n=-\infty}^\infty G(i \nu_n) G(i \Omega_m - i \nu_n) \gamma^{pp}(i \nu_n, i \Omega_m - i \nu_n),\\ 
        P^{ph}(i \Omega_m) &=  \frac{1}{\beta}  \sum_{n=-\infty}^\infty G(i \nu_n) G(i \Omega_m + i \nu_n) \gamma^{ph}(i \nu_n, i \Omega_m + i \nu_n).
    \end{aligned}
\end{equation}
\end{widetext}
Here $G$ is a single-particle Green's function and $\gamma^{\{ph,pp\}}$ are
three-point (Hedin) vertex functions in their respective channels.
For simplicity, we have suppressed orbital and momentum degrees of freedom, but
our algorithm generalizes straightforwardly. Indeed,
including orbital and momentum indices requires computing sums of
the form \eqref{eq:polppph} repeatedly, so it is crucial to have an efficient
algorithm for this step.
The computational cost of direct evaluation of \eqref{eq:polppph} scales as $\OO{\nmax^2}$ in the 
Matsubara frequency cutoff $\nmax$, with only an $\OO{1/\nmax}$ error convergence rate.

By the change of variables $\nu_n \gets -\nu_n$, we can express $P^{pp}$ and $P^{ph}$
in the general form
\begin{equation} \label{eq:polgen}
    P(i \Omega_m) = \frac{1}{\beta} \sum_n F(i \nu_n) G(i \Omega_m - i \nu_n) \gamma(i \nu_n, i \Omega_m - i \nu_n),
\end{equation}
where $F$ and $G$ are single-particle Green's functions, $\gamma$ is a
particle-particle vertex, and we
have suppressed the infinite limits of summation. The polarization $P$ itself
will be represented by its
bosonic DLR expansion. 
By definition \cite{Krien_2021,vanloon23}, $\gamma(i \nu_m, i \nu_n) \to 1$ as $m, n \to \infty$, so we separate out the
constant part explicitly and write the remainder as a DLR expansion \eqref{eq:dlr2d}.
While the validity of this expansion relies on that of the spectral
representation \eqref{eq:specrep_mf} which was derived only for correlation
functions, our numerical evidence presented in Sec.~\ref{sec:results} and that
of Ref.~\onlinecite{wallerberger21}
suggests it is also valid for the non-constant part of
the vertex. A theoretical analysis of this question is an important topic of
future research.

Substituting this decomposition of $\gamma$ into \eqref{eq:polgen} yields
\begin{equation}
    P = P^{(0)} + P^{(1)} + P^{(2)} + P^{(3)} + P^{(4)},
\end{equation}
with
\begin{equation}
    P^{(0)}(i \Omega_m) = \frac{1}{\beta} \sum_n F(i \nu_n) G(i \Omega_m - i \nu_n)
\end{equation}
and
\begin{equation}
    \begin{multlined}
    P^{(\alpha)}(i \Omega_m) \\ = \frac{1}{\beta} \sum_n F(i \nu_n) G(i \Omega_m - i \nu_n) \gamma^{(\alpha)}(i \nu_n, i \Omega_m - i \nu_n),
    \end{multlined}
\end{equation}
for $\alpha = 1,\ldots,4$, with each $\gamma^{(\alpha)}$ corresponding to a term in
the DLR expansion \eqref{eq:dlr2d}:
\begin{equation}
    \begin{multlined}
    \gamma^{(\alpha)}(i \nu_n, i \Omega_m - i \nu_n) \\ =
    \begin{cases}
        \sum_{k,l=1}^r K_k(i\nu_n) K_l(i \Omega_m - i \nu_n)      \wh{\gamma}_{1kl},    &   \alpha = 1,  \\
        \sum_{k,l=1}^r K_k(i \Omega_m - i \nu_n) K_{Bl}(i \Omega_m)  \wh{\gamma}_{2kl}, &   \alpha = 2,  \\
        \sum_{k,l=1}^r K_k(i \nu_n) K_{Bl}(i \Omega_m)               \wh{\gamma}_{3kl}, &   \alpha = 3,  \\
        \delta_{m 0} \sum_{k=1}^r K_k(i \nu_n) \wh{\gamma}_{4kk},         &   \alpha = 4.
    \end{cases}
\end{multlined}
\end{equation}
Here, we have introduced the shorthand $K_k(i \nu_n) = K(i \nu_n, \omega_k)$,
and similarly for $K_B$.

$P^{(0)}$ can be computed straightforwardly by Fourier transformation to imaginary time; we have
\begin{equation}
    P^{(0)}(\tau) = F(\tau) G(\tau),
\end{equation}
which is readily computed by forming the DLR expansions of $F$ and $G$ in the
Matsubara frequency domain and
evaluating them on the imaginary time DLR grid. We can then form the DLR
expansion of the result, and evaluate it back on the Matsubara frequency
DLR grid. The cost of this process is dominated by that of forming and evaluating
the DLR expansions, which scales as $\OO{r^2}$.

The other terms are not simple convolutions, and therefore cannot be computed
straightforwardly by Fourier analysis. However, the DLR expansion enables a
reduction to the convolutional case by separation of variables. We begin with
$P^{(1)}$ which, using the shorthand
\begin{equation}
    \begin{aligned}
        \wt{F}_k(i \nu_n) &= F(i \nu_n) K_k(i \nu_n), \\
        \wt{G}_k(i \nu_n) &= G(i \nu_n) K_k(i \nu_n)
    \end{aligned}
\end{equation}
takes the compact form
\begin{equation}
    P^{(1)}(i \Omega_m) = \frac{1}{\beta} \sum_{k,l=1}^r \wh{\gamma}_{1kl} \sum_n \wt{F}_k(i \nu_n) \wt{G}_l(i \Omega_m - i \nu_n),
\end{equation}
so
\begin{equation}
    P^{(1)}(\tau) = \sum_{k,l=1}^r \wh{\gamma}_{1kl} \wt{F}_k(\tau) \wt{G}_l(\tau).
\end{equation}
Thus, $P^{(1)}(i \Omega_m)$ can be computed by the following procedure:
\begin{enumerate}
    \item Form the DLR expansions of $\wt{F}_k$ and $\wt{G}_l$ in the Matsubara frequency domain, at an $\OO{r^2}$
cost for each of the $2r$ functions, or an $\OO{r^3}$ cost in total. 
    \item Evaluate the expansions at the $r$ imaginary time DLR nodes and take their products, at an
$\OO{r^3}$ cost.
    \item Perform the sum over $k$ and $l$ at each DLR imaginary
time node, at an $\OO{r^3}$ cost.
    \item Form the DLR expansion of the result, at an $\OO{r^2}$ cost.
\end{enumerate}
The total cost is $\OO{r^3}$.
To compute $P^{(2)}$, we write
\begin{equation}
    \begin{multlined}
    P^{(2)}(i \Omega_m) \\ = \frac{1}{\beta} \sum_{k,l=1}^r K_{Bl}(i \Omega_m) \wh{\gamma}_{2kl} \sum_n F(i \nu_n) \wt{G}_k(i \Omega_m - i \nu_n),
    \end{multlined}
\end{equation}
and use the following procedure:
\begin{enumerate}
    \item For each $k$, compute the convolution by a product in imaginary time, as above.
    \item For each $l$, compute the sum over $k$, and transform the result to the imaginary
frequency DLR nodes by DLR expansion.
    \item For each $l$, multiply the result by $K_{Bl}(i \Omega_m)$, and sum over $l$.
\end{enumerate}
The total cost is again $\OO{r^3}$. A similar procedure applies to the remaining terms
\begin{equation}
    \begin{multlined}
    P^{(3)}(i \Omega_m) \\ = \frac{1}{\beta} \sum_{k,l=1}^r K_{Bl}(i \Omega_m) \wh{\gamma}_{3kl} \sum_n \wt{F}_k(i \nu_n) G(i \Omega_m - i \nu_n).
    \end{multlined}
\end{equation}
and
\begin{equation}
    P^{(4)}(i \Omega_m) = \frac{\delta_{m 0}}{\beta} \sum_{k=1}^r \wh{\gamma}_{4kk} \sum_n \wt{F}_k(i \nu_n) G(i \Omega_m - i \nu_n).
\end{equation}

One further technical remark is necessary. Products in the $\tau$ domain in general increase the frequency
content of the result: for example, $K(\tau,\omega_1) K(\tau,\omega_2) =
\frac{K(0,\omega_1) K(0, \omega_2)}{K(0, \omega_1 + \omega_2)} K(\tau, \omega_1
+ \omega_2)$, i.e., the spectral widths add. As a consequence, even if $P$ obeys
a DLR cutoff $\Lambda$, intermediate functions obtained during the course of our
algorithm, such as $\wt{F}_k(\tau) \wt{G}_l(\tau)$, might
not be resolved on the DLR
grid obtained using the DLR cutoff $\Lambda$. However, the frequency
content can at most double, so this issue is remedied by
discretizing $\tau$ in the algorithm above by a DLR grid with cutoff $2
\Lambda$, generating a grid of $r' > r$ nodes. In particular, a DLR
expansion should be evaluated in the $\tau$ domain at these $r'$
nodes; a DLR expansion obtained by interpolation in the $\tau$ domain should
have $r'$ DLR coefficients; and such an expansion should be evaluated back
at only the $r$ nodes in the bosonic Matsubara frequency grid. This
modification does not affect the computational complexity of the algorithm. 

This method generalizes straightforwardly to other diagrammatic expressions in which two legs of $\gamma$ are closed by a pair of propagators. The equation of motion for the self-energy in its Hedin form $\Sigma \sim GW\gamma$ \cite{Krien_2021}, for example, involves such a contraction, with a fermionic propagator $G$ and a bosonic propagator $W$.

In Appendix \ref{app:matsumres}, we propose an alternative algorithm to compute
the polarization \eqref{eq:polppph} via the residue theorem, exploiting the meromorphic structure of the
DLR expansion extended to the complex plane. The computational complexity of this algorithm is
also $\OO{r^3}$, but it is somewhat more complicated and expensive. However,
we include it since the technique is quite general, and may be useful in
other diagrammatic calculations.

\section{Numerical results} \label{sec:results}

\subsection{Hubbard atom}

To demonstrate our approach, we first consider the atomic limit of an isolated yet interacting spinful fermion with Hamiltonian
\begin{align}
	\mathcal{H} = U n_{\uparrow} n_{\downarrow} - \mu (n_{\uparrow} +  n_{\downarrow}) \,,
	\label{eq:atomic_limit}
\end{align}
where $n_{\sigma}$ is the occupation number operator for spin $\sigma \in \{ \uparrow, \downarrow \}$. We
choose the chemical potential $\mu = U/2$ so that the system is half-filled.
The correlation functions can be computed analytically in this case,
yet they contain non-trivial structure,
for example the formation of a local magnetic moment beyond simple
perturbation theory \cite{Wentzell_2020, Thunstroem_2018}. The Hubbard atom
therefore provides a convenient and useful test of the validity and efficiency
of our approach.

Defining $\Pi(i \nu_m, i \nu_n) = G(i\nu_m) G(i\nu_n)$ for the
atomic Green's function \mbox{$G(i\nu_n) = (i\nu_n - \tfrac{U^2}{4 i \nu_n})^{-1}$},
the three-point correlators in the singlet ($si$), charge ($ch$) and spin ($sp$)
channels (corresponding to certain linear combinations of spin components in the
particle-particle and particle-hole channels, respectively
\cite{Rohringer_2018}) are given by
\begin{widetext}
\begin{equation}
\begin{aligned}
	\chi^{si}(i\nu_m, i\nu_n) &=
        \paren{\frac{U^2}{2 \nu_m \nu_n} + 2} \Pi(i\nu_m, i\nu_n) -\delta_{\nu_m + \nu_n,0} \beta U n_F\paren{\frac{U}{2}} \paren{\frac{U^2}{4 \nu_m^2} + 1} \Pi(i\nu_m, -i\nu_m), \\
	\chi^{\alpha}(i\nu_m, i\nu_n) &= 
        \paren{\frac{U^2}{4 \nu_m \nu_n} - 1} \Pi(i\nu_m, i\nu_n) + \delta_{\nu_m - \nu_n,0} \paren{\frac{\beta U^\alpha}{2} n_F\paren{\frac{U^\alpha}{2}} \paren{\frac{U^2}{4 \nu_m^2} + 1} \Pi(i\nu_m, i\nu_m) + \delta_{\alpha,ch} \beta G(i\nu_m)},
\end{aligned}
\end{equation}
\end{widetext}
for $n_F(\omega) = 1 / (1 + e^{\beta \omega})$ the Fermi-Dirac function, and $\alpha \in \{ch, sp\}$. Here $U^{ch} = U$ and $U^{sp} = -U$. These three-point correlation
functions obey the representation \eqref{eq:specrep_mf} and its corresponding DLR expansion \eqref{eq:dlr2d}. One
can also obtain a closed expression for the respective Hedin vertices,
\begin{equation}
\begin{multlined}
	\gamma^{si}(i\nu_m, i\nu_n) \hfill \\= \frac{2 + \frac{U^2}{2 \nu_m \nu_n} - \delta_{\nu_m + \nu_n,0} \beta U n_F\paren{\frac{U}{2}} \paren{1 + \frac{U^2}{4 \nu_m^2}}}{2 - \delta_{\nu_m + \nu_n,0} \beta U n_F\paren{\frac{U}{2}}}, \\
	\gamma^{\alpha}(i\nu_m, i\nu_n) \hfill \\= \frac{2 - \frac{U^2}{2 \nu_m \nu_n} - \delta_{\nu_m - \nu_n,0} \beta U^\alpha n_F\paren{\frac{U^\alpha}{2}} \paren{1 + \frac{U^2}{4 \nu_m^2}}}{2 - \delta_{\nu_m - \nu_n,0} \beta U^\alpha n_F\paren{\frac{U^\alpha}{2}}},
\end{multlined}
\end{equation}
for $\alpha \in \{ch, sp\}$, by amputating the associated bosonic and fermionic propagators from $\chi$.
Finally, the polarization functions in the three channels are given by 
\begin{equation}
\begin{aligned}
	P^{si}(i \Omega_n) &= \delta_{n 0} \frac{\beta n_F(U/2)}{2 U \beta n_F(U/2) - 4}, \\  
	P^{\alpha}(i \Omega_n) &= \delta_{n 0} \frac{\beta n_F(U^\alpha/2)}{U^\alpha \beta n_F(U^\alpha/2) - 2},
\end{aligned}
\end{equation}
revealing that only long-lived (static) bosonic fluctuations survive in the
atomic limit.

\begin{figure*}
    \includegraphics[width=0.32\textwidth]{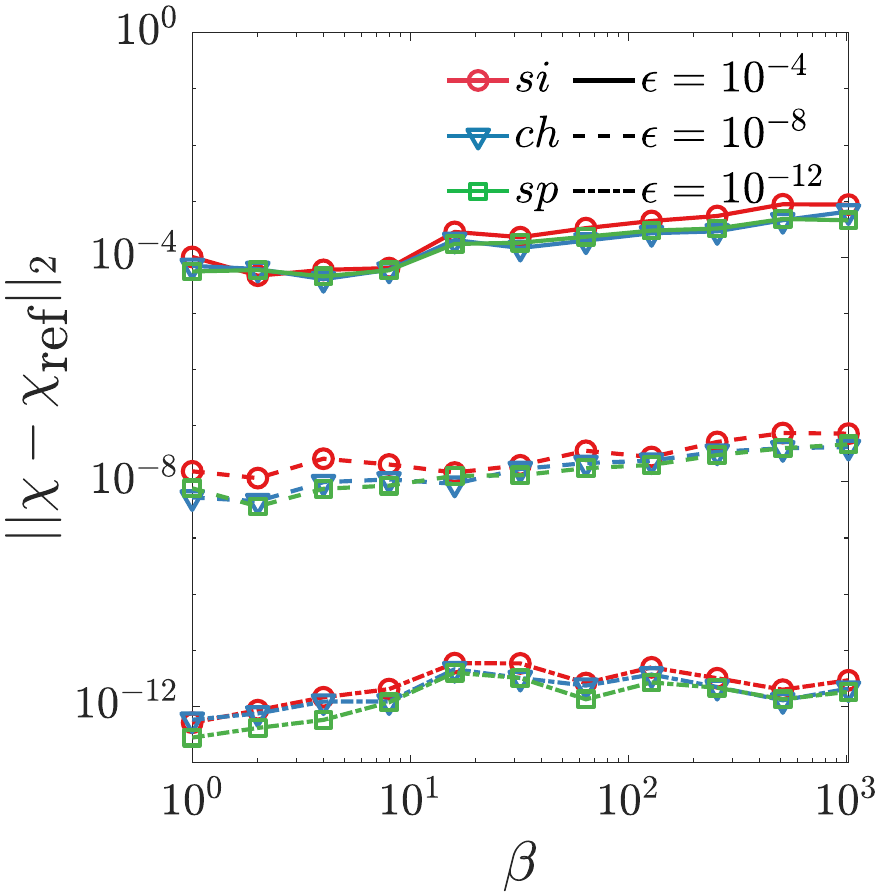}
    \hfill
    \includegraphics[width=0.32\textwidth]{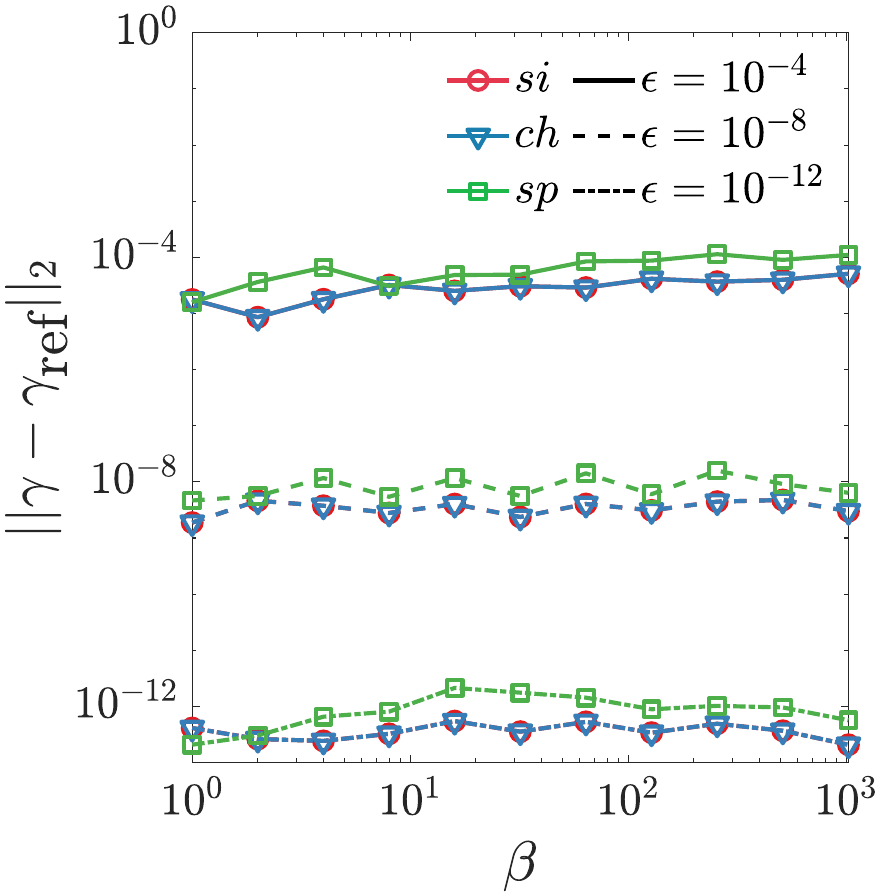}
    \hfill
    \includegraphics[width=0.32\textwidth]{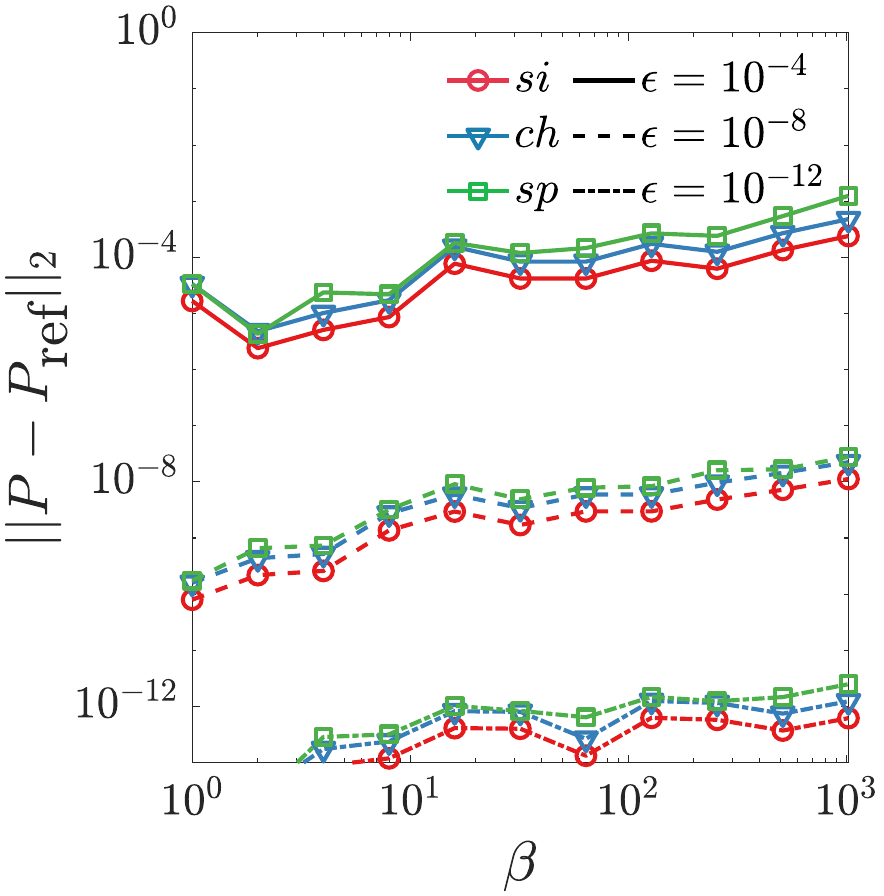}
    \caption{$l^2(i \nu_n)$ error, compared with an exact reference, of the DLR expansion as a function of $\beta$ and the truncation tolerance
    $\epsilon$, for the correlation, vertex, and polarization
    functions (computed using the method of Sec. \ref{sec:matsum}) of the Hubbard atom example.}
    \label{fig:hub_errvbeta}
\end{figure*}

Given these analytic expressions, we can test the accuracy of the DLR expansions
of $\chi^{\alpha}$ and $\gamma^{\alpha}$, $\alpha \in \{si, ch, sp\}$, obtained by the interpolation
method described in Sec. \ref{sec:dlrinterp}, as well as the DLR expansions of
$P^{\alpha}$, obtained from the DLR
expansions of $G$ and the $\gamma^{\alpha}$ by the Matsubara
summation method described in Sec. \ref{sec:matsum}. We measure the error in the
$l^2$ norms given by \eqref{eq:l2norm} (with $\nu_n$ replaced by
$\Omega_n$) for the $P^\alpha$, and
\begin{equation} \label{eq:l2norm2d}
    \begin{aligned}
        \norm{\chi}_2^2 &= \frac{1}{\beta^2} \int_0^\beta \int_0^\beta d\tau_1 d\tau_2 \, \abs{\chi(\tau_1,\tau_2)}^2 \\
        &= \frac{1}{\beta^4} \sum_{m,n=-\infty}^\infty \abs{\chi(i \nu_m, i \nu_n)}^2.
    \end{aligned}
\end{equation}
for $\chi^\alpha$ and $\gamma^\alpha$, which we compute in the Matsubara frequency domain by
evaluation on a large grid. We fix $\Lambda = \beta$. Our results are shown in Fig.~\ref{fig:hub_errvbeta},
where we measure the error as a function of $\beta$ for $\epsilon = 10^{-4}$,
$10^{-8}$, and $10^{-12}$. We observe errors close to these specified tolerances
up to $\beta = 1000$. The number of degrees of freedom $R$ required by our grid
representation for each $\beta = \Lambda$ and $\epsilon$ can be deduced from
Fig. \ref{fig:rankvlambda}. For example, for $\epsilon = 10^{-8}$ and $\Lambda =
1024$, we require $1185$ degrees of freedom, or roughly $18.5$ KB of memory.

\subsection{Single impurity Anderson model}

For a proof-of-concept without an analytical solution, and involving polarization functions with dynamics
at non-zero frequency, we couple the interacting orbital to a
collection of non-interacting electronic levels: 
\begin{align}
	\mathcal{H} &= U n_{\uparrow} n_{\downarrow} - \mu (n_{\uparrow} +  n_{\downarrow}) \notag \\ &+ \sum_{i \sigma} \paren{V_i c^{\dagger}_{i \sigma} d^{\phantom{\dagger}}_{\sigma} + \text{h.c.}} + \sum_{i \sigma} \epsilon_{i} c^{\dagger}_{i \sigma} c^{\phantom{\dagger}}_{i \sigma}.
\end{align}
Here $d^{(\dagger)}_{\sigma}$ and $c^{(\dagger)}_{i \sigma}$ are the creation
and annihilation operators for the impurity and bath sites, respectively,
with the latter having energies $\epsilon_i$.
We choose the hybridization $V_i = 1$ as our unit of energy and consider a
particle-hole symmetric setting with $\mu = U/2$ and three
levels $\epsilon_i \in \{ 0, \pm U \}$, such that the bath has
bandwidth $2U$. To obtain a reference solution, we use the \emph{pomerol} exact
diagonalization library \cite{pomerol}, which directly outputs the required
one-, two-, and three-point correlators on a dense Matsubara frequency grid. The
polarization functions and vertices are computed using the relations derived
in Secs. IIB and IIF of Ref.~\onlinecite{Krien_2020}.
We obtain data on the DLR grid by subsampling from the dense Matsubara frequency
grid. We use $\beta = 20$ and $U = 5$.

\begin{figure*}
    \includegraphics[width=0.32\textwidth]{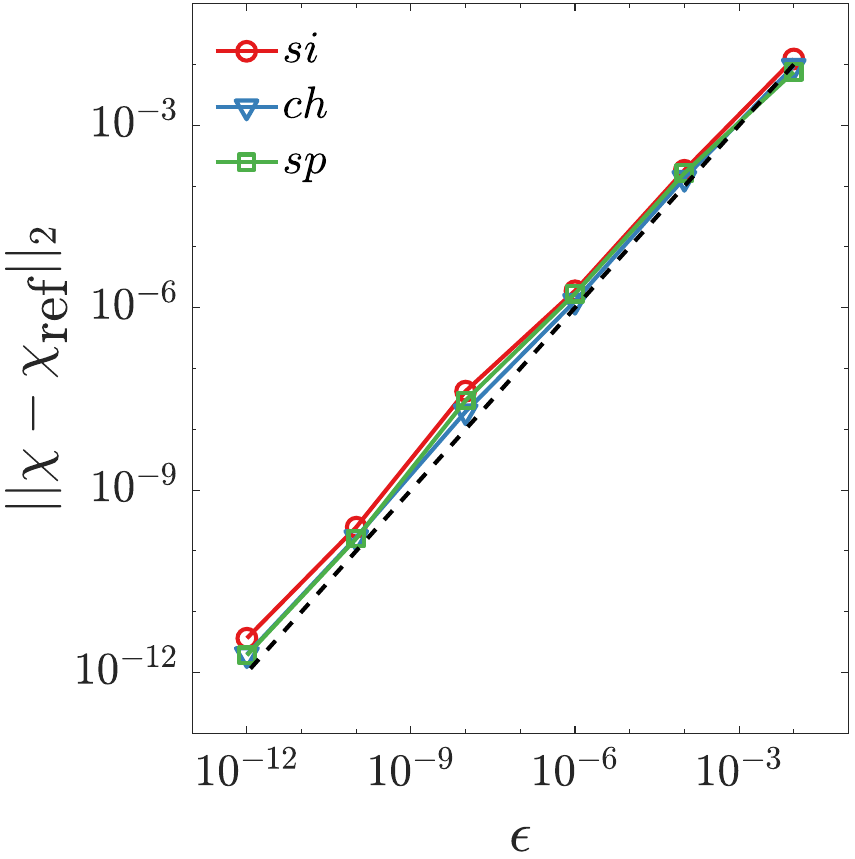}
    \hfill
    \includegraphics[width=0.32\textwidth]{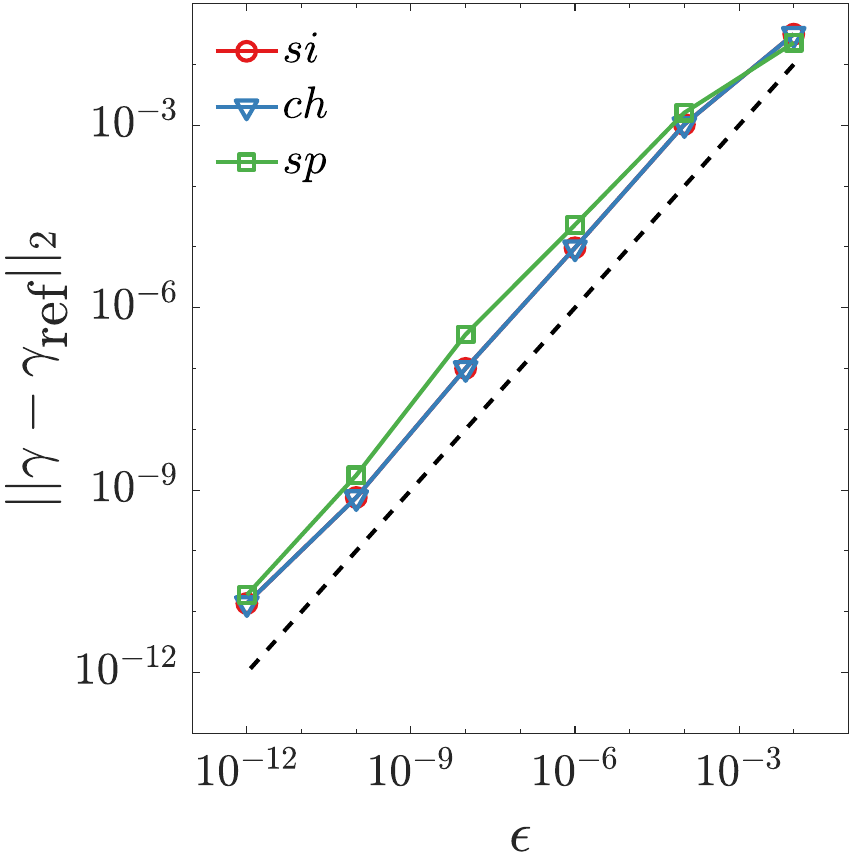}
    \hfill
    \includegraphics[width=0.32\textwidth]{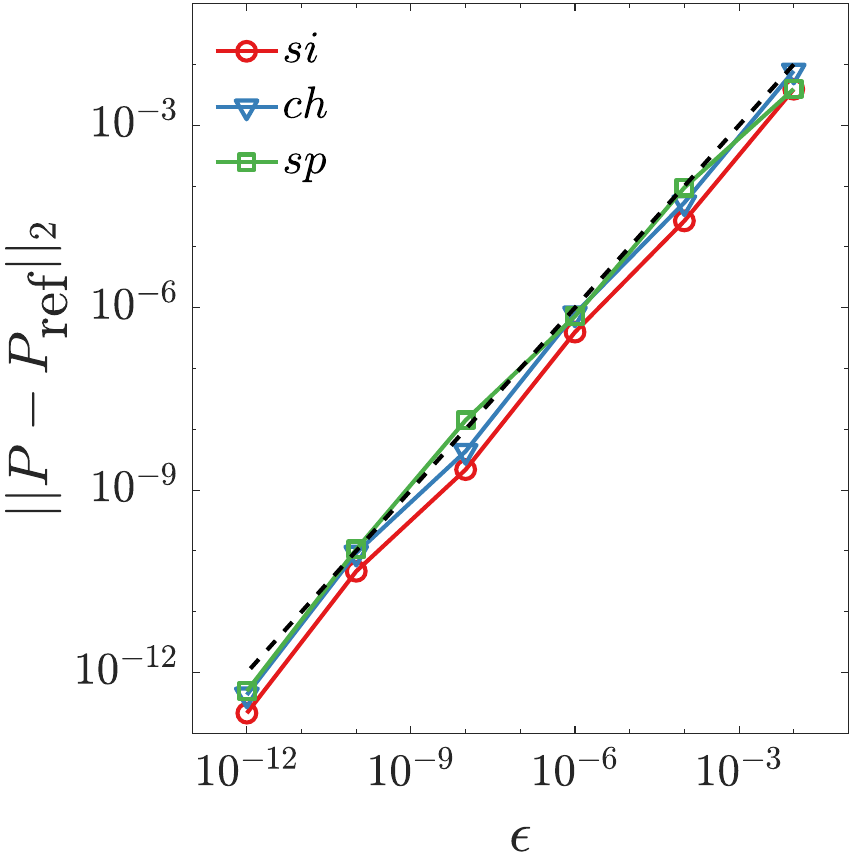}
    \caption{$l^2(i \nu_n)$ error, compared with a reference computed by exact diagonalization, of the DLR expansion as a function of the truncation tolerance
    $\epsilon$ for the correlation, vertex, and polarization
    functions (computed using the method of Sec. \ref{sec:matsum}) of the single impurity Anderson model example. Error
    $\epsilon$ is also indicated as a dashed line, and we observe errors close to this specified
    tolerance.}
    \label{fig:siam_errveps}
\end{figure*}

As for the Hubbard atom case, we measure the $l^2$ error of the DLR expansions
of $\chi^\alpha$ and $\gamma^\alpha$, obtained via interpolation, and 
$P^\alpha$, obtained via our Matsubara summation algorithm. Results are given in
Fig.~\ref{fig:siam_errveps} as a function of the truncation tolerance
$\epsilon$, using $\Lambda = 300$. We
observe convergence down to machine precision, with the error typically
within one significant digit of the specified tolerance. A careful analysis of the
relationship between the truncation tolerance and the guaranteed error is a
topic of our future research. 

\section{Discussion}

We have presented a representation and associated algorithms for the efficient manipulation of three-point correlation and vertex
functions. Immediate applications include methods based on three-point vertices, such as the TRILEX and dual-TRILEX extensions of DMFT \cite{Ayral_2016, Stepanov_2019}, as well as the inclusion of vertex corrections within the (ab-initio) $GW$ framework \cite{Aryasetiawan_1998}. This preliminary work suggests a number of research questions and directions for generalization and improvement.

An important question is the validity of the spectral representation 
\eqref{eq:specrep_mf} and its $n$-point generalization
\cite{shinaoka18} for vertex functions, on which our approach relies. Although
such spectral representations have been studied in detail for correlation
functions \cite{kugler21,dirnbock24}, to the authors' knowledge a proof of their validity
for $n$-particle vertex functions with $n > 1$ (the existence of a spectral representation for the dynamic part of the self-energy is demonstrated in Ref.~\cite{Luttinger_1961}) has not been presented beyond the perturbation limit, with the numerical results here and in previous work \cite{wallerberger21} for specific cases comprising the
principal evidence.

Although the generalization is straightforward for certain classes of diagrams,
the applicability of our Matsubara frequency summation methods to other
diagrammatic expressions must also be explored further. We expect broadly that
the availability of a simple, explicit representation creates possibilities for
efficient algorithms in many physically-relevant settings. 

Another important follow-up is the generalization of our approach to four-point
functions. In principle, this is straightforward, with
the essential ingredients in place (spectral representations are described in
Refs.~\onlinecite{shinaoka18} and \onlinecite{kugler21}). However,
at sufficiently low temperatures, the computational effort and memory of factorizing and
storing the matrix $\Phi^f$ (scaling as $\OO{\log^9(\Lambda)}$ and
$\OO{\log^6(\Lambda)}$, respectively) as well as applying its pseudoinverse to
obtain the DLR expansion coefficients (scaling as $\OO{\log^6(\Lambda)}$) could
become prohibitive. Ref.~\onlinecite{wallerberger21} circumvents this by using an
iterative solver for fitting, with an
$\OO{\log^5(\Lambda)}$-scaling matrix application scheme, but the cost over
many iterations remains substantial, and a closer analysis of the required number of
iterations, and its dependence on $\Lambda$, is needed. Algorithms with further reduced complexity might exist,
for example via an iterative solver and a fast algorithm to apply $\Phi$,
direct compression of its pseudoinverse, or a tensor network approximation of
the expansion coefficients \cite{shinaoka20}.

\acknowledgments

We thank Fabian Kugler for helpful discussions. HURS acknowledge funding from the European Research Council (ERC) under the European Union’s Horizon 2020 research and innovation programme (Grant agreement No.\ 854843-FASTCORR). KC was supported by the National Natural Science Foundation of China under Grants No. 12047503.
The Flatiron Institute is a division of the Simons Foundation.

\onecolumngrid

\appendix

\section{Matsubara frequency summation via residue calculus} \label{app:matsumres}

\begin{figure*}
    \includegraphics[width=0.32\textwidth]{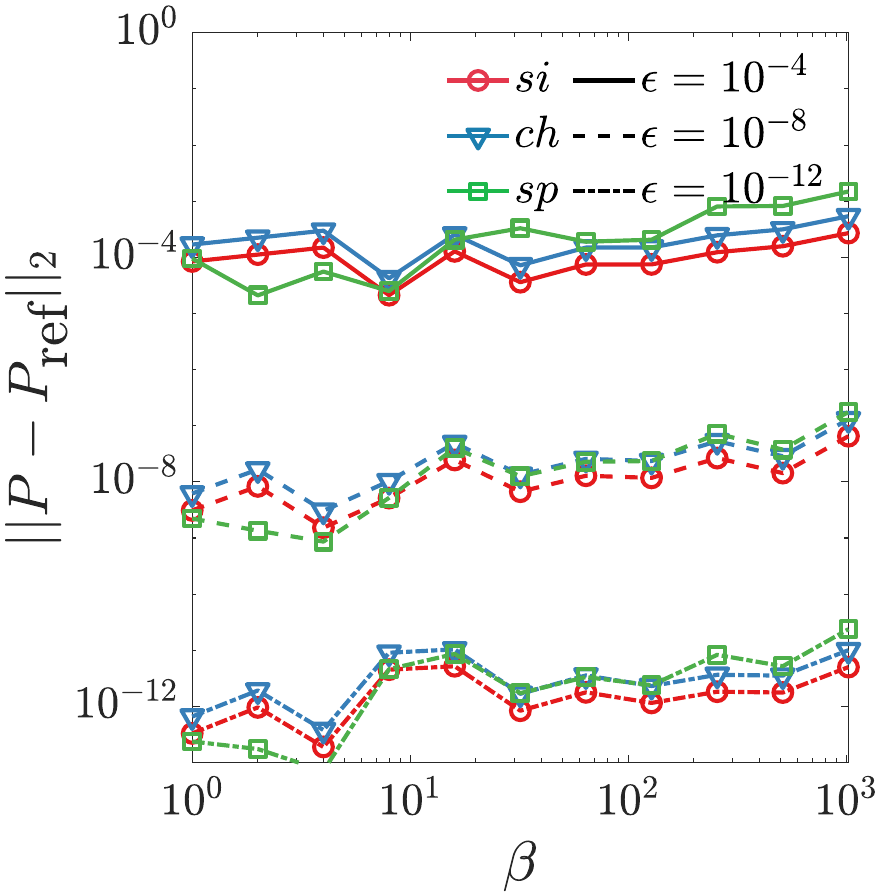}
    \hspace{0.15\textwidth}
    \includegraphics[width=0.32\textwidth]{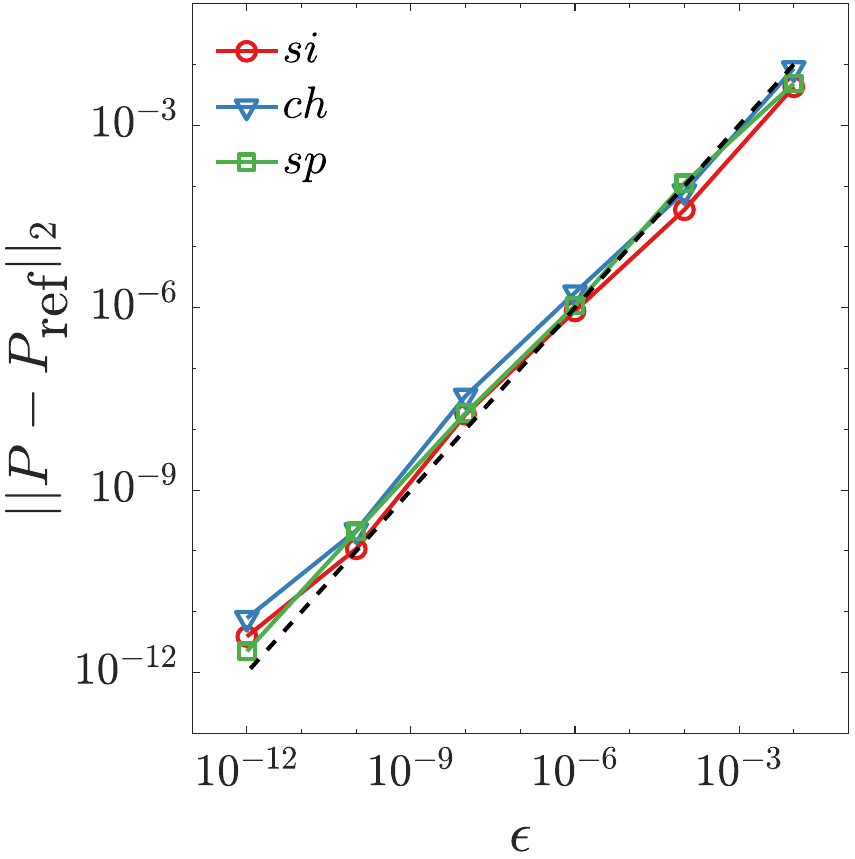}
    \caption{$l^2(i \nu_n)$ error of the polarization functions computed using the residue calculus
    algorithm of Appendix \ref{app:matsumres}, shown as a function of $\beta$ and the truncation tolerance
    $\epsilon$ for the Hubbard atom example (left), and as a function of the
    truncation tolerance for the single impurity Anderson model example (right).
    These can be compared with the third panels of Figs. \ref{fig:hub_errvbeta}
    and \ref{fig:siam_errveps}, respectively, which use the Matsubara summation
    algorithm described in Sec.~\ref{sec:matsum}.}
    \label{fig:polbyres}
\end{figure*}

In this Appendix we describe an alternative $\OO{r^3}$-scaling algorithm for
computing the polarization \eqref{eq:polgen}, based on the meromorphic structure
of the DLR expansion and residue calculus. Although it is slower 
than the algorithm presented in Sec.~\ref{sec:matsum}, we include it because of
its apparent generality. Fig. \ref{fig:polbyres} repeats the third panels
of Figs. \ref{fig:hub_errvbeta} and \ref{fig:siam_errveps}, with the algorithm
of Sec.~\ref{sec:matsum} replaced by the residue calculus algorithm described below.

We again note that $P(i \Omega_m)$ need only be computed on the $r$ bosonic DLR
nodes. We can compute the contribution
$P^{(0)}$ of the constant part of the vertex as before, so in the remainder of this section we ignore it,
making the replacement $\gamma \gets \gamma - 1$ in the definition of $P$.

We next treat the zero-frequency case,
\begin{equation}
    \begin{aligned}
    P(i \Omega_0) &= \frac{1}{\beta} \sum_n F(i \nu_n) G(-i \nu_n) \gamma(i \nu_n, -i \nu_n) \\
    &= \frac{1}{\beta} \sum_n H(i \nu_n) = H(\tau = 0),
    \end{aligned}
\end{equation}
where the second equality defines the summand, and the third uses
the Fourier transform to imaginary time.
This can be obtained by
forming the DLR expansion of $H$ in the Matsubara frequency domain and
evaluating it in the imaginary time domain at $\tau = 0$, at an $\OO{r}$ cost by
precomputing the corresponding linear functional.

We finally consider $m \neq 0$, and begin by replacing $F(i \nu_n)$, $G(i \nu_n)$, and $\gamma(i \nu_m, i \nu_n)$
by their DLR expansions. 
Then for each fixed $\Omega_m$, $m \neq 0$, we define the function
\begin{equation}
    \psi_m(z) = F(z) G(i \Omega_m - z) \gamma(z, i \Omega_m - z),
\end{equation}
where $F$, $G$, and $\gamma$ have been extended to the complex plane by
replacing the $i \nu_n$ arguments in their DLR expansions by the complex
variable $z \in \mathbb{C}$:
\begin{equation}
\begin{aligned}
    F(z) &= \sum_{k=1}^r \frac{\wh{f}_k}{z - \omega_k}, \\
    G(i \Omega_m - z) &= \sum_{k=1}^r \frac{\wh{g}_k}{i \Omega_m - z - \omega_k}, \\
    \gamma(z, i \Omega_m - z) &= (\gamma^{(1)} + \gamma^{(2)} + \gamma^{(3)})(z, i \Omega_m - z),
\end{aligned}
\end{equation}
with
\begin{equation}
    \gamma^{(\alpha)}(z, i \Omega_m - z) =
    \begin{cases}
    \displaystyle
    \sum_{k,l=1}^r \frac{\wh{\gamma}_{1kl}}{(z - \omega_k)(i \Omega_m - z - \omega_l)}, & \alpha = 1,\\
    \displaystyle
    \sum_{k,l=1}^r \frac{\wh{\gamma}_{2kl} \tanh(\beta \omega_l / 2)}{(i \Omega_m - z - \omega_k)(i \Omega_m - \omega_l)}, & \alpha = 2,\\
    \displaystyle
    \sum_{k,l=1}^r \frac{\wh{\gamma}_{3kl} \tanh(\beta \omega_l / 2)}{(z - \omega_k)(i \Omega_m - \omega_l)}, & \alpha = 3.
    \end{cases} 
\end{equation}
Note that we ignore the singular term of the DLR expansion of $\gamma$,
since it only contributes to $P(i \Omega_0)$.
Evidently, $\psi$ is meromorphic, and we have
\begin{equation} \label{eq:polbyres}
    P(i \Omega_m) = \frac{1}{\beta} \sum_n \psi_m(i \nu_n) = \sum_{j=1}^{n_r} \mathrm{Res}(\psi_m n_F, z_j).
\end{equation}
Here the $z_j$ are
the $n_r$ poles of $\psi_m$. This expression can be obtained by applying the
Cauchy residue theorem with two contours enclosing the positive and negative
Matsubara frequencies, respectively, which are the poles of $n_F$. Thus, to
compute $P(i \Omega_m)$ at the non-zero DLR nodes $i \Omega_{m_k}$, we must
determine the pole locations $z_j$ of $\psi_{m_k}$, compute the corresponding
residues of $\psi_{m_k} n_F$, and apply \eqref{eq:polbyres}.
To organize the calculation, we define $\psi_m^{(\alpha)}(z) = F(z) G(i \Omega_m
- z) \gamma^{(\alpha)}(z, i \Omega_m - z)$, compute \eqref{eq:polbyres} with
$\psi_m$ replaced by $\psi_m^{(\alpha)}$ for the terms $\alpha = 1, 2, 3$ in turn,
and sum the results. The pole locations and residues can be determined
explicitly from the DLR expansions of $F$, $G$, and $\gamma^{(\alpha)}$.

The details for each term are given below, and we obtain an algorithm with an
$\OO{r^3}$ computational complexity in total. The efficiency of the algorithm
stems from two properties: (1) the total number of residues is only $\OO{r}$,
and (2) the expressions for the residues can be computed in a nested manner. By
contrast, computing the polarization by substituting in the DLR expansions of
$F$, $G$, and $\gamma$, and carrying out the resulting tensor contraction has a
computational complexity of $\OO{r^5}$.

\begin{remark}
    One might be concerned by the complex analytic structure of
    the underlying functions $F$, $G$, and $\gamma$; for instance,
    that branch cuts, or poles close to the imaginary axis, would
    render \eqref{eq:polbyres} invalid. However, since the DLR expansions of these
    functions are correct within a controlled error in the Matusbara
    frequency domain, for the purpose of computing $P(i \Omega_m)$
    using residue calculus they can be replaced by these expansions, which have a simple
    meromorphic structure.
\end{remark}

\begin{remark}
    Rather than performing the residue calculation manually for each diagram topology,
    it might be possible to adapt existing ideas from the literature to automate this process.
    For example, the algorithmic Matsubara integration method \cite{taheridehkordi19,elazab22,burke23,assi24} uses
    residue calculus along with symbolic algebra to compute the Matsubara sums
    arising in bare perturbative expansions, using a representation of the
    non-interacting Green's function by a simple pole.
    A similar idea is presented in Ref.~\onlinecite{Jaksha_2021}, but using
    an imaginary time formulation rather than residue calculus.
    Also related is the method of Ref.~\onlinecite{kaye23_diagrams}, in which a
    pole expansion is used to obtain efficient and explicit methods for
    evaluating arbitrary imaginary time Feynman diagrams appearing in bold
    diagrammatic expansions, in a manner which can be automated without symbolic
    algebra \cite{huang24}.
\end{remark}

\subsection{First term}

We have
\begin{equation}
    \psi_m^{(1)}(z) = \sum_{pqkl} \frac{\wh{f}_p \wh{g}_q \wh{\gamma}_{1kl}}{(z-\omega_p)(z-\omega_k)(i \Omega_m - z -
\omega_q)(i \Omega_m - z - \omega_l)}.
\end{equation}
Here, we have suppressed the limits of summation over $r$ terms.
$\psi_m^{(1)}(z)$ has double poles at $z = \omega_j$ and $z = i \Omega_m -
\omega_j$, for $j \in \{1,\ldots,r\}$.

We begin with the residue at $z = \omega_j$. Only the terms with $p =
j$ or $k = j$ contribute to the residue. We first consider the contribution of the terms
with $p = j$, $k \neq j$, and $k = j$, $p \neq j$, for which the pole is simple:
\begin{equation}
    \begin{multlined}
    \Res^{(1)}(\psi_m^{(1)} n_F, \omega_j) 
\\ = n_F(\omega_j) \paren{\sum_q \frac{\wh{g}_q}{i \Omega_m - \omega_j - \omega_q}} \paren{\wh{f}_j \sum_l \frac{1}{i \Omega_m - \omega_j - \omega_l} \sum_{k \neq j}\frac{\wh{\gamma}_{kl}}{\omega_j-\omega_k} + \sum_l \frac{\wh{\gamma}_{jl}}{i \Omega_m - \omega_j - \omega_l} \sum_{p \neq j}\frac{\wh{f}_p}{\omega_j-\omega_p}}.
    \end{multlined}
\end{equation}
This expression must be evaluated for each of the non-zero DLR nodes $i
\Omega_{m_k}$, and the $r$ residues $\omega_j$. Carried out in the nested manner
indicated, the cost of each such evaluation is $\OO{r}$, yielding an overall
$\OO{r^3}$ cost.
We next consider the contribution of the terms with $p = k = j$, for which there
is a double pole:
\begin{equation}
\begin{aligned}
    \Res^{(2)}(\psi_m^{(1)} n_F, \omega_j) &= \wh{f}_j \lim_{z \to \omega_j} \frac{d}{dz} \sum_{ql} \frac{\wh{g}_q \wh{\gamma}_{jl} n_F(z)}{(i \Omega_m - z - \omega_q)(i \Omega_m - z - \omega_l)} \\
    &= \begin{multlined}[t] \wh{f}_j \left( n_F'(\omega_j) \sum_q \frac{\wh{g}_q}{i \Omega_m - \omega_j - \omega_q} \sum_l \frac{\wh{\gamma}_{jl} }{i \Omega_m - \omega_j - \omega_l} \right. \\
        + n_F(\omega_j) \sum_q \frac{\wh{g}_q}{(i \Omega_m - \omega_j - \omega_q)^2} \sum_l \frac{\wh{\gamma}_{jl} }{i \Omega_m - \omega_j - \omega_l} \\
        \left. + n_F(\omega_j) \sum_q \frac{\wh{g}_q}{i \Omega_m - \omega_j - \omega_q} \sum_l \frac{\wh{\gamma}_{jl} }{(i \Omega_m - \omega_j - \omega_l)^2}\right).
    \end{multlined}
\end{aligned}
\end{equation}
The total cost of evaluating each of these $r$ residues for all DLR nodes in the
nested manner indicated is again $\OO{r^3}$.
We then have $\Res(\psi_m^{(1)} n_F, \omega_j) = \Res^{(1)}(\psi_m^{(1)} n_F, \omega_j) + \Res^{(2)}(\psi_m^{(1)} n_F, \omega_j)$.

For the residue at $z = i \Omega_m - \omega_j$, the
contribution of the terms with $q = j$, $l \neq j$, and $l = j$, $q \neq j$ is
\begin{equation}
    \begin{multlined}
    \Res^{(1)}(\psi_m^{(1)} n_F, i \Omega_m - \omega_j) 
\\ = -n_F(- \omega_j) \paren{\sum_p \frac{\wh{f}_p}{i \Omega_m - \omega_j - \omega_p}} \paren{\wh{g}_j \sum_k \frac{1}{i \Omega_m - \omega_j - \omega_k} \sum_{l \neq j}\frac{\wh{\gamma}_{kl}}{\omega_j - \omega_l} + \sum_k \frac{\wh{\gamma}_{kj}}{i \Omega_m - \omega_j - \omega_k} \sum_{q \neq j}\frac{\wh{g}_q}{\omega_j - \omega_q}}.
    \end{multlined}
\end{equation}
Here we have used that $n_F(i \Omega_m - \omega_j) = -n_F(-\omega_j)$.
The contribution of the terms with $q = l = j$ is
\begin{equation}
\begin{aligned}
    \Res^{(2)}(\psi_m^{(1)} n_F, i \Omega_m - \omega_j) &= \wh{g}_j \lim_{z \to i\Omega_m - \omega_j} \frac{d}{dz} \sum_{pk} \frac{\wh{f}_p \wh{\gamma}_{kj} n_F(z)}{(z - \omega_p)(z - \omega_k)} \\
    &= \begin{multlined}[t] \wh{g}_j \left( n_F'(-\omega_j) \sum_p \frac{\wh{f}_p}{i \Omega_m - \omega_j - \omega_p} \sum_k \frac{\wh{\gamma}_{kj} }{i \Omega_m - \omega_j - \omega_k} \right. \\
        - n_F(-\omega_j) \sum_p \frac{\wh{f}_p}{(i \Omega_m - \omega_j - \omega_p)^2} \sum_k \frac{\wh{\gamma}_{kj}}{i \Omega_m - \omega_j - \omega_k} \\
        \left. - n_F(-\omega_j) \sum_p \frac{\wh{f}_p}{i \Omega_m - \omega_j - \omega_p} \sum_k \frac{\wh{\gamma}_{kj}}{(i \Omega_m - \omega_j - \omega_k)^2}\right),
    \end{multlined}
\end{aligned}
\end{equation}
and $\Res(\psi_m^{(1)} n_F, i \Omega_m - \omega_j) = \Res^{(1)}(\psi_m^{(1)}
n_F, i \Omega_m - \omega_j) + \Res^{(2)}(\psi_m^{(1)} n_F, i \Omega_m -
\omega_j)$. The total cost of computing this expression is again $\OO{r^3}$, as
it will be for the remaining contributions discussed below. This completes the calculation
of $\Res(\psi_m^{(1)} n_F, z_j)$ for each of the residues $z_j$.

\subsection{Second term}

We have
\begin{equation}
    \psi_m^{(2)}(z) = \sum_{pqkl} \frac{\wh{f}_p \wh{g}_q \wh{\gamma}_{2kl}}{(z - \omega_p)(i \Omega_m - z - \omega_q)(i \Omega_m - z - \omega_k)(i \Omega_m - \omega_l)}
    = \sum_{pqk} \frac{\wh{f}_p \wh{g}_q \wh{\Gamma}_{2km}}{(z - \omega_p)(i \Omega_m - z - \omega_q)(i \Omega_m - z - \omega_k)},
\end{equation}
with
\begin{equation} \label{eq:biglambda}
    \wh{\Gamma}_{2km} = \sum_{l} \frac{\wh{\gamma}_{2kl}}{i \Omega_m - \omega_l}.
\end{equation}
$\psi_m^{(2)}(z)$ has simple poles at $z = \omega_j$, and double poles
at $z = i \Omega_m - \omega_j$, for $j \in \{1,\ldots,r\}$.

We first consider the residues at $z = \omega_j$. The only terms which contribute
to the residue are those with $p = j$:
\begin{equation}
\Res(\psi_m^{(2)} n_F, \omega_j) \\ = \wh{f}_j n_F(\omega_j) \sum_q \frac{\wh{g}_q}{i\Omega_m - \omega_j - \omega_q} \sum_{k}\frac{\wh{\Gamma}_{2km}}{i \Omega_m - \omega_j - \omega_k}.
\end{equation}

Let us next consider the pole at $z = i \Omega_m - \omega_j$. We
first consider the contribution of the terms with $q = j$, $k \neq j$ and
$k = j$, $q \neq j$, for which the pole is simple:
\begin{equation}
\begin{aligned}
\Res^{(1)}(\psi_m^{(2)} n_F, i \Omega_m - \omega_j) &= -\wh{g}_j n_F(- \omega_j) \sum_{p,k \neq j} \frac{\wh{f}_p \wh{\Gamma}_{2km}}{(i \Omega_m - \omega_j - \omega_p)(\omega_j - \omega_k)} -
n_F(- \omega_j) \sum_{p,q \neq j} \frac{\wh{f}_p \wh{g}_q \wh{\Gamma}_{2jm}}{(i \Omega_m - \omega_j - \omega_p)(\omega_j - \omega_q)} \\
&= -n_F(- \omega_j) \paren{\sum_p \frac{\wh{f}_p}{i \Omega_m - \omega_j - \omega_p}} \paren{\wh{g}_j \paren{\sum_{k \neq j} \frac{\wh{\Gamma}_{2km}}{\omega_j-\omega_k}} 
+ \wh{\Gamma}_{2jm} \sum_{q \neq j} \frac{\wh{g}_q}{\omega_j-\omega_q}}.
\end{aligned}
\end{equation}
We next consider the contribution of the terms with $q = k = j$, for which there
is a double pole:
\begin{equation}
    \Res^{(2)}(\psi_m^{(2)} n_F, i \Omega_m - \omega_j) = g_j \wh{\Gamma}_{2jm} \lim_{z \to i \Omega_m - \omega_j} \frac{d}{dz} \sum_{p} \frac{\wh{f}_p n_F(z)}{z-\omega_p} 
    = g_j \wh{\Gamma}_{2jm} \paren{\sum_{p} \frac{n_F'(-\omega_j) \wh{f}_p}{i \Omega_m - \omega_j - \omega_p} - \frac{n_F(-\omega_j) \wh{f}_p}{(i \Omega_m - \omega_j - \omega_p)^2}}.
\end{equation}
We then have
\begin{equation}
    \Res(\psi_m^{(2)} n_F, i \Omega_m - \omega_j) \\ = \Res^{(1)}(\psi_m^{(2)} n_F, i \Omega_m - \omega_j) + \Res^{(2)}(\psi_m^{(2)} n_F, i \Omega_m - \omega_j).
\end{equation}

\subsection{Third term}

We have
\begin{equation}
\psi_m^{(3)}(z) = \sum_{pqk} \frac{\wh{f}_p \wh{g}_q \wh{\Gamma}_{3km}}{(z-\omega_p)(z-\omega_k)(i
\Omega_m - z - \omega_q)},
\end{equation}
with $\wh{\Gamma}_{3km}$ defined as in \eqref{eq:biglambda}.
$\psi_m^{(3)}(z)$ has double poles at $z = \omega_j$ and simple poles at $z = i
\Omega_m - \omega_j$, for $j \in \{1,\ldots,r\}$.
We begin with the residue at $z = \omega_j$. The contribution of the terms with $p =
j$, $k \neq j$, and $k = j$, $p \neq j$ is
\begin{equation}
    \Res^{(1)}(\psi_m^{(3)} n_F, \omega_j) 
= n_F(\omega_j) \paren{\sum_q \frac{\wh{g}_q}{i \Omega_m - \omega_j - \omega_q}} \paren{\wh{f}_j \sum_{k \neq j} \frac{\wh{\Gamma}_{3km}}{\omega_j-\omega_k} + \wh{\Gamma}_{3jm} \sum_{p \neq j} \frac{\wh{f}_p}{\omega_j-\omega_p}}.
\end{equation}
The contribution of the terms with $p = k = j$ is
\begin{equation}
    \Res^{(2)}(\psi_m^{(3)} n_F, \omega_j) = \wh{f}_j \wh{\Gamma}_{3jm} \lim_{z \to \omega_j} \frac{d}{dz} \sum_q \frac{\wh{g}_q n_F(z)}{i \Omega_m - z - \omega_q} 
    = \wh{f}_j \wh{\Gamma}_{3jm} \paren{\sum_{q} \frac{n_F'(\omega_j) \wh{g}_q}{i \Omega_m - \omega_j - \omega_q} + \frac{n_F(\omega_j) \wh{g}_q}{(i \Omega_m - \omega_j - \omega_q)^2}},
\end{equation}
and we have $\Res(\psi_m^{(3)} n_F, \omega_j) = \Res^{(1)}(\psi_m^{(3)} n_F, \omega_j) + \Res^{(2)}(\psi_m^{(3)} n_F, \omega_j)$.

For the residue at $z = i \Omega_m - \omega_j$, the only contributing terms are
those with $q = j$:
\begin{equation}
    \Res(\psi_m^{(3)} n_F, i \Omega_m - \omega_j) = -\wh{g}_j n_F(- \omega_j) \sum_{p} \frac{\wh{f}_p}{i \Omega_m - \omega_j - \omega_p} \sum_{k} \frac{\wh{\Gamma}_{3km}}{i \Omega_m - \omega_j - \omega_k}.
\end{equation}

\twocolumngrid

\bibliographystyle{apsrev4-2}
\bibliography{dlrnd}

\end{document}